\documentclass[runningheads]{llncs}
\usepackage{graphicx}
\usepackage{color,soul}
\usepackage{algorithmic}
\usepackage{algorithm}
\usepackage[table,xcdraw,usenames,dvipsnames]{xcolor}
\usepackage{subcaption}
\usepackage{amssymb,mathtools}
\usepackage{enumitem}
\let\llncssubparagraph\subparagraph
\let\subparagraph\paragraph
\usepackage[compact]{titlesec}
\let\subparagraph\llncssubparagraph

\usepackage[compact]{titlesec}  

\begin{document}
\title{A Reinforcement Learning Approach for Re-allocating Drone Swarm Services}

\author{Balsam Alkouz\inst{1}\orcidID{0000-0001-7938-4438} \and
Athman Bouguettaya\inst{1}\orcidID{0000-0003-1254-8092}}

\institute{University of Sydney, NSW, Australia
\email{\{balsam.alkouz,athman.bouguettaya\}@sydney.edu.au}}

\titlerunning{An RL Approach for Re-allocating Drone Swarm Services}
\maketitle              
\begin{abstract}
We propose a novel framework for the re-allocation of drone swarms for delivery services known as Swarm-based Drone-as-a-Service (SDaaS). The re-allocation framework ensures maximum profit to drone swarm providers while meeting the time requirement of service consumers. The constraints in the delivery environment (e.g., limited recharging pads) are taken into consideration. We utilize reinforcement learning (RL) to select the best allocation and scheduling of drone swarms given a set of requests from multiple consumers. We conduct a set of experiments to evaluate and compare the efficiency of the proposed approach considering the provider's profit and run-time efficiency.

\end{abstract}
\section{Introduction}
\emph{Swarm-based Drone-as-a-Service (SDaaS)} is a concept that describes services offered by swarms of drones \cite{alkouz2020swarm}. The SDaaS notion is an augmentation on the Drone-as-a-Service (DaaS) concept that describes services offered by single drones \cite{hamdi2021drone}. It offers added capabilities to cover services a single drone is not capable of achieving. Examples of these services include search and rescue \cite{cardona2019robot}, sky shows and entertainment \cite{waibel2017drone}, and delivery of goods \cite{alkouz2020formation}. Our focus is on the use of drone swarms in delivery. An increasing dependency on drone delivery is perceived especially during pandemics, as they are contact-less and fast. Therefore, robust and effective deliveries of multiple/heavier packages are needed. Such deliveries are only possible using a swarm of drones as flight regulations only allow the use of small drones (payload$<$2.5 kg) to deliver in the city\footnote{https://www.faa.gov/uas/advanced\_operations/package\_delivery\_drone}. In addition, swarms of drones in delivery are capable of covering longer trips by distributing the payload over several drones decreasing the rate of battery consumption \cite{alkouz2020swarm}.
Swarm-based drone deliveries operating in a city are assumed to be flying within line of sight segments in a skyway network \cite{lee2021package}. The skyway network nodes are assumed to be building rooftops equipped with recharging pads that a swarm may land on to extend its flight range \cite{shahzaad2020game}. We formally define an SDaaS as a swarm carrying packages and travelling in a skyway segment frome node A to node B. The composition of optimal segments between a source node and a destination node would result in an optimal composite SDaaS service. An SDaaS service maps to the key components of service computing, i.e. functional and non-functional attributes \cite{shahzaad2021robust}. The function of an SDaaS is the successful delivery of packages by a swarm between two nodes. The non-functional attributes or the Quality of Services (QoS) include the delivery time, cost, etc.

Three main steps are involved in a successful SDaaS delivery. First, an optimal swarm members allocation approach is essential to serve multiple consumers requests in a day. Second, an optimal path composition method is required to optimize the QoS. Third, a failure-recovery solution is necessary in case of uncertainties. In this paper, we focus on the first step, with respect to the composition, to optimally allocate swarms to consumers requests from a provider point of view. The last step, i.e. failure recovery, is the future extension of this work. An optimal allocation is key in assuring that a provider owned drones are optimally \textit{utilized} and \textit{re-utilized} within a day. Therefore, fulfilling as many consumers requests as possible and increasing a provider profit.\looseness=-1

There are several challenges in the swarms allocation problem. First, a provider owns a \textit{limited set of drones} that needs to be utilized maximally. Second, the delivery time of consumers requests may overlap as they need to be delivered within \textit{strict time windows}. Hence, requests that maximize the providers profit need to be allocated. Third, the requests need to be served in a way that optimizes the \textit{re-utilization} of drones. Hence, within a time window, a swarm may be reused if its round trip time to the first request is smaller than the time window. This problem is challenging since the allocation of any swarm is highly dependent on the availability of other drones because they are \textit{re-allocatable}. In addition, each swarm is bounded by a \textit{Round Trip Time} from the source to the destination and back to the destination. This means that the allocation of any request highly \textit{affects the allocation of other requests in the same time-window and other windows} as the provider owns a limited set of drones.
We propose to allocate \textit{any} available drones to \textit{multiple time-constrained} requests, and \textit{re-utilize} the drones multiple times within a time window to \textit{maximize a providers profit}. We summarize our main contributions as following: \looseness=-1
\begin{itemize}[nosep]
    \item A modified A* congestion-aware algorithm to compose SDaaS services.
    \item An RL SDaaS allocation algorithm to maximize providers profit.
\end{itemize}

\section{Related Work}
A robotic swarm is a set of robots that collectively solve a problem to achieve a common goal. 
In delivery, majority of literature refer to swarms of drones as multiple single independent drones managed to deliver multiple independent deliveries \cite{kuru2019analysis}. However, we refer to a swarm as a set of drones carrying multiple packages for a single delivery operation. In this regard, a sequential and parallel delivery services composition using a swarm of drones was proposed \cite{alkouz2020swarm}. 
While drone swarms in delivery represent a major advancement, developing \textit{swarm allocation} methods is essential to unlock their full potential and obtain teamwork benefit \cite{gigliotta2018equal}.\looseness=-1


Multi-Robot Task Allocation (MRTA) addresses the assignment of set of tasks to a set of robots \cite{elfakharany2020towards}. The robots need to be optimally allocated to tasks to optimize the overall team performance \cite{khamis2015multi}. Multi Robot Task Scheduling (MRTS) deals with the scheduling of the tasks to minimize the overall cost, make it be: time, money, or energy. Most multi-robot systems deal with MRTA and MRTS as two different steps. However, the decoupling of these steps leads to partial observability and lack of full insights \cite{elfakharany2020towards}. In addition, to the best of our knowledge, most work done in MRTA does not deal with the \textit{multiple re-allocations} of the robots in a \textit{time-constrained} environment. Hence, we propose to couple the MRTA and MTRS problems and deal with multiple re-allocations of drone swarms in a time-constrained environment using a \textit{service-oriented approach}.\looseness=-1

The service paradigm is a key enabler of drone deliveries in a skyway network. It ensures congruent and effective provisioning of drone-based deliveries \cite{shahzaad2021resilient}. 
Previous works discuss the optimal composition of services, i.e. composing the best path from the source to the destination \cite{alkouz2020swarm}. 
In a different application, a reinforcement learning approach to compose moving WiFi hotspot services was proposed \cite{gharineiat2021deep}. Majority of the existing work uses deep reinforcement learning for services composition and not allocation \cite{wang2010adaptive}. Hence, this work is the first that deals with the \textit{re-allocation} of SDaaS services to optimize the QoS. 
This work takes into consideration the optimal \textit{SDaaS composition} and challenges due to the simultaneous use of the skyway network by multiple swarms.\looseness=-1


\section{Swarm-based Drone-as-a-Service Model}
\begin{figure}
\centering
\includegraphics[width=\linewidth]{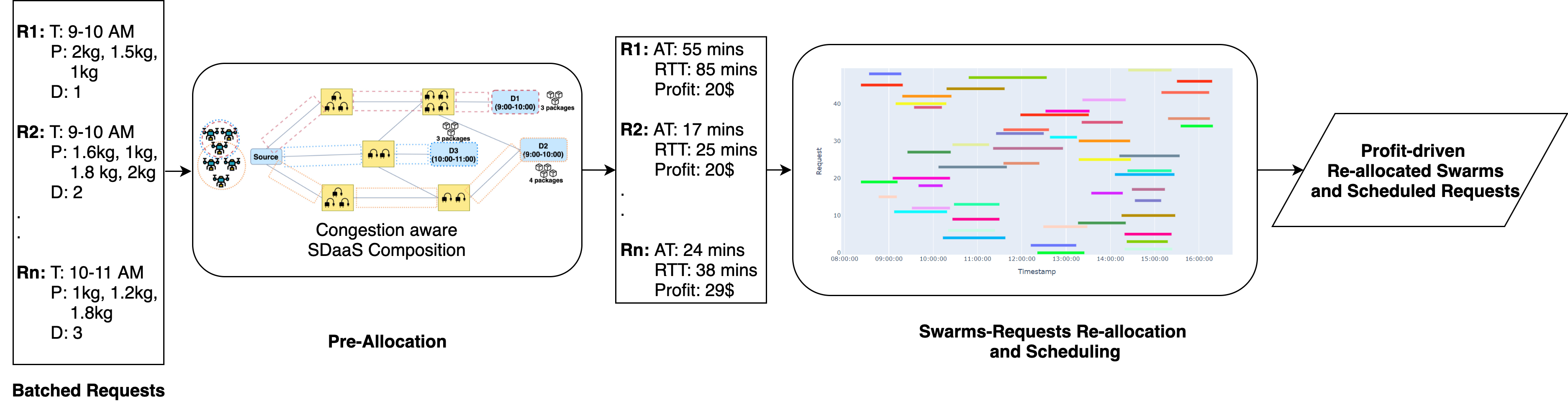}
\caption{SDaaS Members Re-allocation Framework}
\label{fig:framework}
\end{figure}

In this section, we present a swarm-based drone delivery service model. We abstract a swarm carrying packages and travelling in a skyway segment between two nodes as a service (Fig. \ref{fig:framework}).\\ 
\textbf{Definition 1: Swarm-based Drone-as-a-Service (SDaaS).} An SDaaS is defined as a set of drones, carrying packages and travelling in a skyway segment. It is represented as a tuple of $<SDaaS\_id, S, F>$, where
\begin{itemize}[nosep]
    \item $SDaaS\_id$ is a unique service identifier
    \item $S$ is the swarm travelling in SDaaS. $S$ consists of $D$ which is the set of drones forming $S$, a tuple of $D$ is presented as $<d_1,d_2,..,d_m>$. $S$ also contains the properties including the current battery levels of every $d$ in $D$ $<b_1,b_2, ..,b_m>$, the payloads every $d$ in $D$ is carrying $<p_1,p_2,..,p_m>$, and the current node $n$ the swarm S is at.
    \item F describes the delivery function of a swarm on a skyway segment between two nodes, A and B. F consists of the segment distance $dist$, travel time $tt$,  charging time $ct$, and waiting time $wt$ when recharging pads are not enough to serve $D$ simultaneously in node B.
\end{itemize}
\textbf{Definition 2: SDaaS Request.} A request is a tuple of $< R\_id,\beta, P, T>$, where
\begin{itemize}[nosep]
    \item $R\_id$ is the request unique identifier.
    \item $\beta$ is the request destination node.
    \item $P$ are the weights of the packages requested, where $P$ is $<p_1,p_2,..,p_m>$.
    \item $T$ is the time window of the expected delivery, it is represented as a tuple of the window start and end times $<st,et>$.
\end{itemize}

\section{SDaaS Members Re-allocation Framework}

The SDaaS members allocation and scheduling framework composes of two main modules. In the first module, the composition of SDaaS services for every received request is performed. The output of the first module is the maximum time taken for the packages to arrive at the destination (AT), the maximum round trip time back to the source (RTT), and the profit if the request is served. In the second module, the AT, RTT, and profit are used to allocate and re-allocate the provider owned drones to the most profitable requests and schedule them in a way that serves as many requests as possible.

\subsection{SDaaS Pre-Allocation}
\label{composition}


The pre-allocation module mainly consists of the SDaaS optimal composition of all requests to their respective destinations. The optimal path that reduces the delivery time is composed. The intermediate nodes contain different numbers of recharging pads. The composition should consider the optimal selection of nodes that would reduce the charging times. In addition, contention may occur at a node if two swarms serving different requests take the same path at a time causing \textit{congestion} \cite{alkouz2021provider}. We assume that the weight of the packages do not exceed a drones payload capacity. We also assume that a drone may carry a single package at a time. The swarm is assumed to serve one request in a single trip. Therefore, the size of a swarm, serving a request, is equivalent to the number of packages in the request. A request is assumed to have a maximum capacity of $m$ packages. The goal of this module is to compute the maximum time a swarm would take to serve a request (AT) and come back to the source (RTT). The RTT is the maximum possible time of a trip with the existence of other swarms in the network at the same time. Hence, the composition is considered congestion-aware. The composed path is an optimal path in terms of delivery time that a swarm may take while considering the probability of having other swarms utilizing the charging pads, i.e. congestion. Hence, we propose a modified congestion-aware A* approach for SDaaS composition. The AT of the packages at the destination and the RTT is key in scheduling the requests to serve as many requests as possible. We assume that the environment is deterministic, i.e. we know the availability of recharging pads considering other providers using the network.


The composition is initiated with a set of swarm drones ($S_D$), fully charged at the source. The swarm is assumed to be static \cite{akram2017security}, i.e. it traverses the network without splitting midway. 
While the drone is not at the destination and back at the source, the algorithm computes the likelihood for the swarm to reach the dest/src nodes using Dijkstra's shortest path without stopping at intermediate nodes. The likelihood of reaching is computed based on the payload of all the drones and the energy consumption rate over the distance travelled. If the swarm is capable of reaching the dest/src directly, it traverses the network and the $RTT$ gets updated with the travel time $tt$. Otherwise, if the swarm is not capable of reaching the dest/src node directly, it selects the optimal neighbouring node. An optimal neighbor is a neighbouring node with the least travel time $tt$ and node time $nt$. The $nt$ is dependant on the number of available recharging pads at a node. The $nt$ composes of the charging times $ct$ and the waiting times $wt$ due to sequential charging in case the number of pads is less than the size of the swarm. We assume that a node may be used by a maximum of two swarms at a time. At every node, we consider the potential of congestion to compute the maximum possible $AT$ and $RTT$. We assume that each drone is occupied by all the other drones owned by the provider $P_D$ if they are less than the maximum swarm size $m$. Otherwise, we assume a station is used by another swarm of size $m$. We compute the node time considering the number of available recharging pads under congestion. When the best neighbour is selected, the swarm traverses to the node and charges fully. The swarm attempts again to reach the dest/src directly. The process continues until the swarm is at the dest/src. The $RTT$ is updated to include the charging time back at the source. The profit is computed using the number of drones utilized to serve a request $S_D$ and the $RTT$ of the trip.\looseness=-1 


\subsection{SDaaS Allocation and Scheduling}
\label{allocation}
The composed services from subsection \ref{composition} are used to allocate drones to the most profitable requests for the provider. There might be instances where aggregated less profitable requests result in a better total profit than few high profit requests. Hence, the allocation and scheduling algorithm needs to maximize the total profit per day. These allocated requests need to be scheduled in the timeline efficiently to serve as many possible requests. The allocation and scheduling should take in consideration the limited number of provider owned drones. At a time $t$ a provider may serve a maximum of $N$ packages at a time. Therefore, we propose a reinforcement learning allocation and scheduling algorithm.\looseness=-1

\subsubsection{Reinforcement Learning based allocation.}
The proposed framework aims to allocate the provider owned drones $P_D$ to the consumers requests $R$ in the best possible manner, i.e. maximize profit. This imposes the maximum utilization of drones and scheduling the request in the best possible timings to be able to re-allocate the drones over and over again. We leverage \textit{Reinforcement Learning (RL)} to find, allocate, and schedule the requests. In RL, an \textit{agent} learns about an \textit{environment's} behaviour through \textit{explorations}. RL is capable of discovering the best set of requests to be allocated and schedule them at the most optimal time to facilitate the re-use of drones. The main reason for our choice of RL is its \textit{ability to discover the “cumulative” optimal set of requests} to be allocated. The RL does that by assigning rewards for every action the agent invokes. In our work, the actions are the service requests and time slots that a swarm can get allocated to. The agent’s role is to pick the next service request and allocation time that would maximize the overall reward. Therefore, the agent should not only consider the current requests to make the selection but also future requests and available drones. The \textit{environment} that the agent interacts with in this solution is designed to be problem specific. The environment checks for requests validity, overlapping allocations, drones’ availability, and time inter-dependencies and permits only valid actions to be taken by the agent.

We define the agent's \textit{actions} as a tuple of request ID and time slot $<R_{id}, AT_w>$. The time slot represents the arrival time within the consumer specified delivery time window $<R_{st}, R_{et}>$. The agent at every step takes an action, i.e. adds a specific request to the environment at a certain time window. The environment checks the validity of allocating the request by looking at the overlapped allocated requests and the availability of the provider owned drones. 
The \textit{state} is updated at every step with the total accumulated profit of the allocated requests. 
We implement a \textit{Q-learning} algorithm that seeks to find the best action to take given the current state \cite{watkins1992q}. 

\section{Experiments}
\label{experiments}

In this section, we evaluate the performance in terms of total profit gained and the execution time of the proposed algorithm. 
A brute force baseline is time and memory extensive and is not feasible as described earlier. Therefore, we compare the proposed RL allocation method to the First Come First Served (FCFS) algorithm \cite{tanenbaum2015modern}.  In the FCFS approach, the first request received gets allocated first. If a request can’t be allocated due to the limited number of drones being occupied at a time window, the request does not get allocated and the next arriving request gets checked and allocated. 

An urban road network dataset from the city of london is used to mimic the arrangement of a skyway network \cite{karduni2016protocol}. The dataset consists of nodes representing intersections and segments connecting those nodes. For the experiments, we extracted a sub-network consisting of 129 connected nodes. Each node is allocated with different number of recharging pads randomly. A source node is then selected and $r$ service requests are generated with different destination nodes. For each request, we synthesize maximum 5 packages payload and a maximum weight of 1.4kg. The drone model is assumed to be the DJI phantom 3. All the power consumption computation is based on this model, the distance travelled, and payload carried. We used the congestion-aware SDaaS composition algorithm to compute the $AT$, $RTT$, and profit for each request given the recharging pads constraints. These requests are assigned to different time windows randomly. Each time window is assumed to be one hour. Hence, the $AT$ of the package should lie within this hour. The experiments were run on 7th Gen Intel® Core™ i7- 7700HQ Processor (2.8 GHz), 16 GB RAM, 64-bit Windows OS PC. \looseness=-1

\begin{figure}[!ht]
\begin{subfigure}{.5\textwidth}
  \centering
  \includegraphics[width=.9\linewidth]{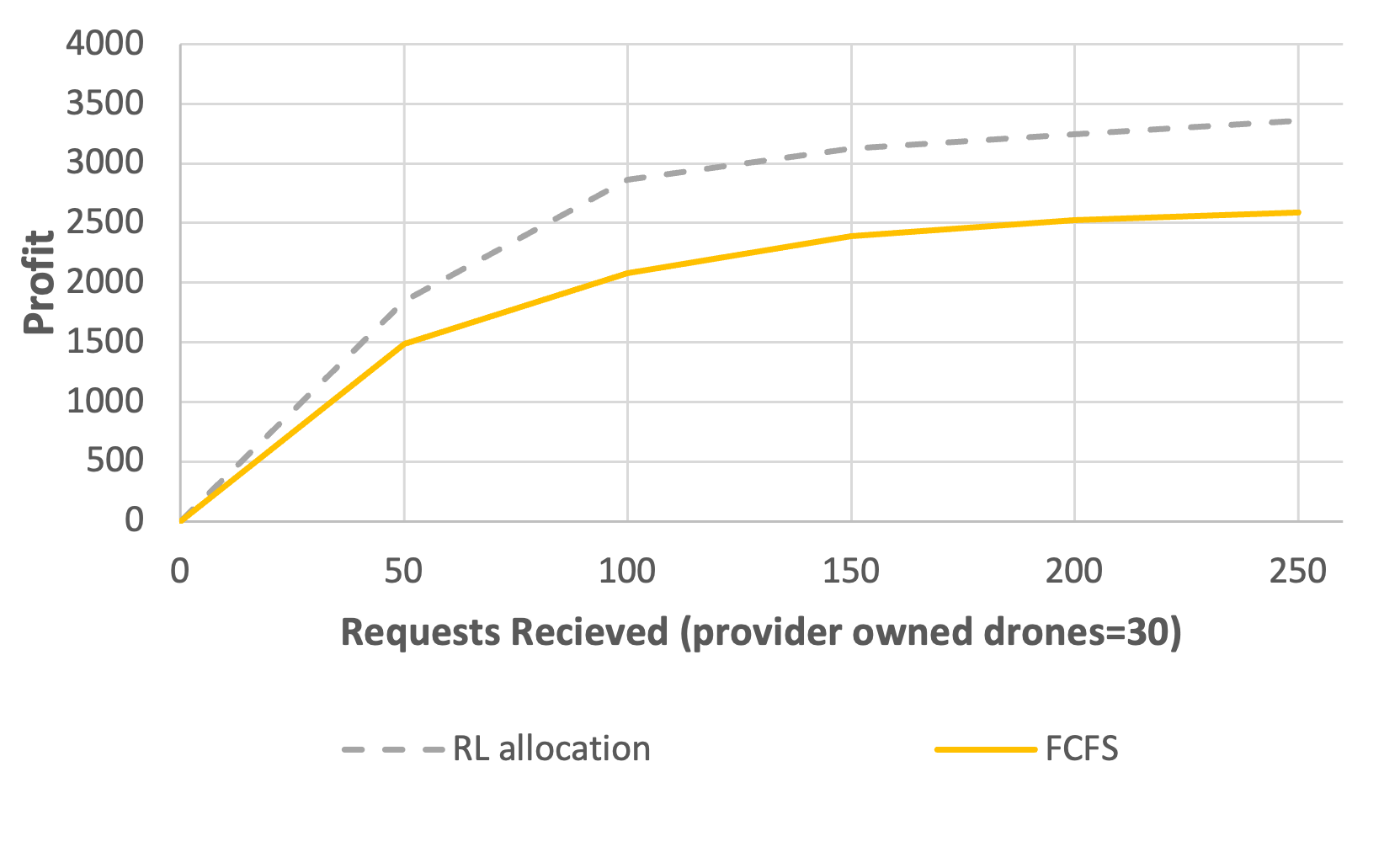}  
  \caption{Profit with varying number of requests}
  \label{fig:sub-first}
\end{subfigure}
\begin{subfigure}{.5\textwidth}
  \centering
  \includegraphics[width=.9\linewidth]{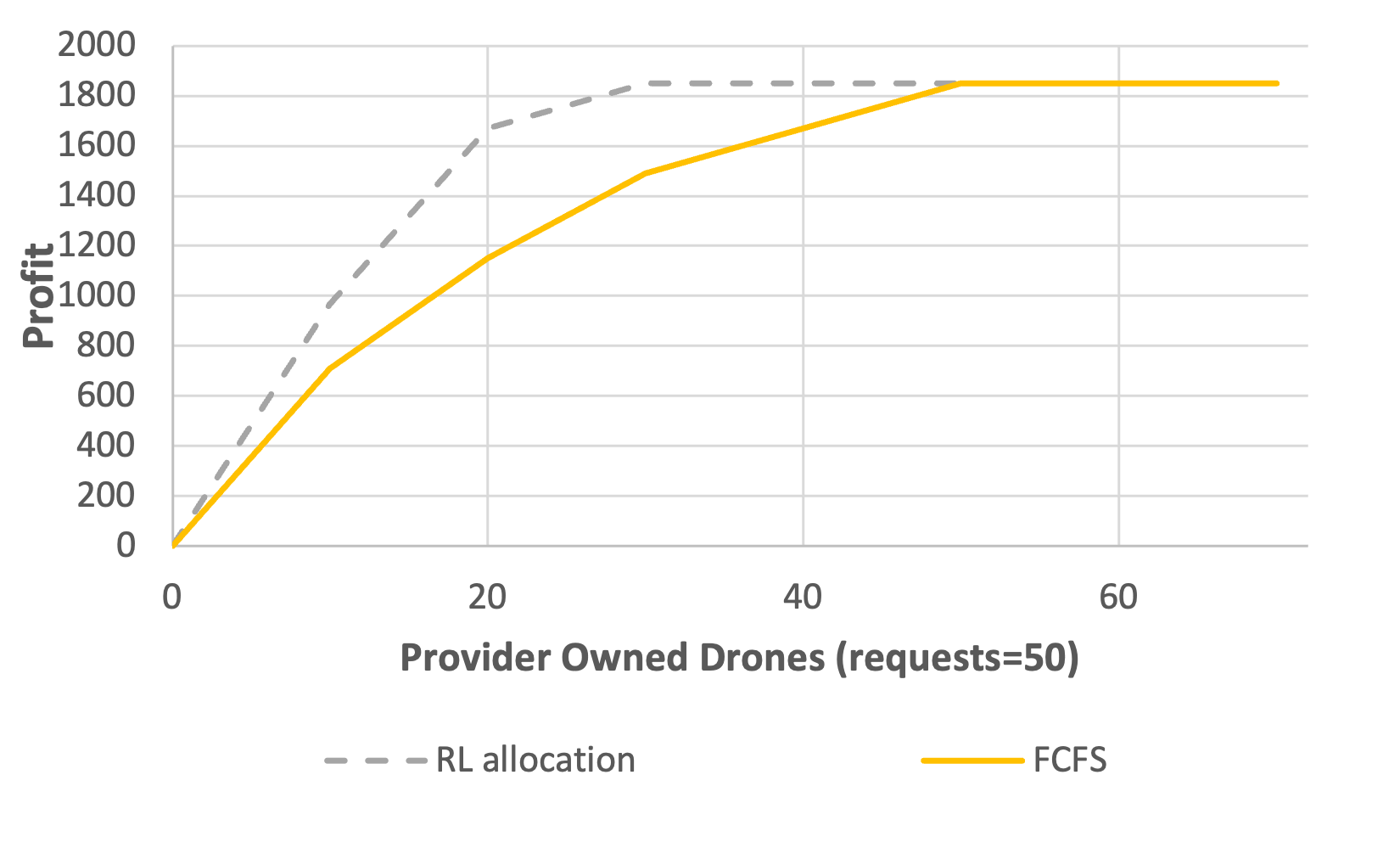}  
  \caption{Profit with varying number of drones}
  \label{fig:sub-second}
\end{subfigure}



\begin{subfigure}{.5\textwidth}
  \centering
  \includegraphics[width=.9\linewidth]{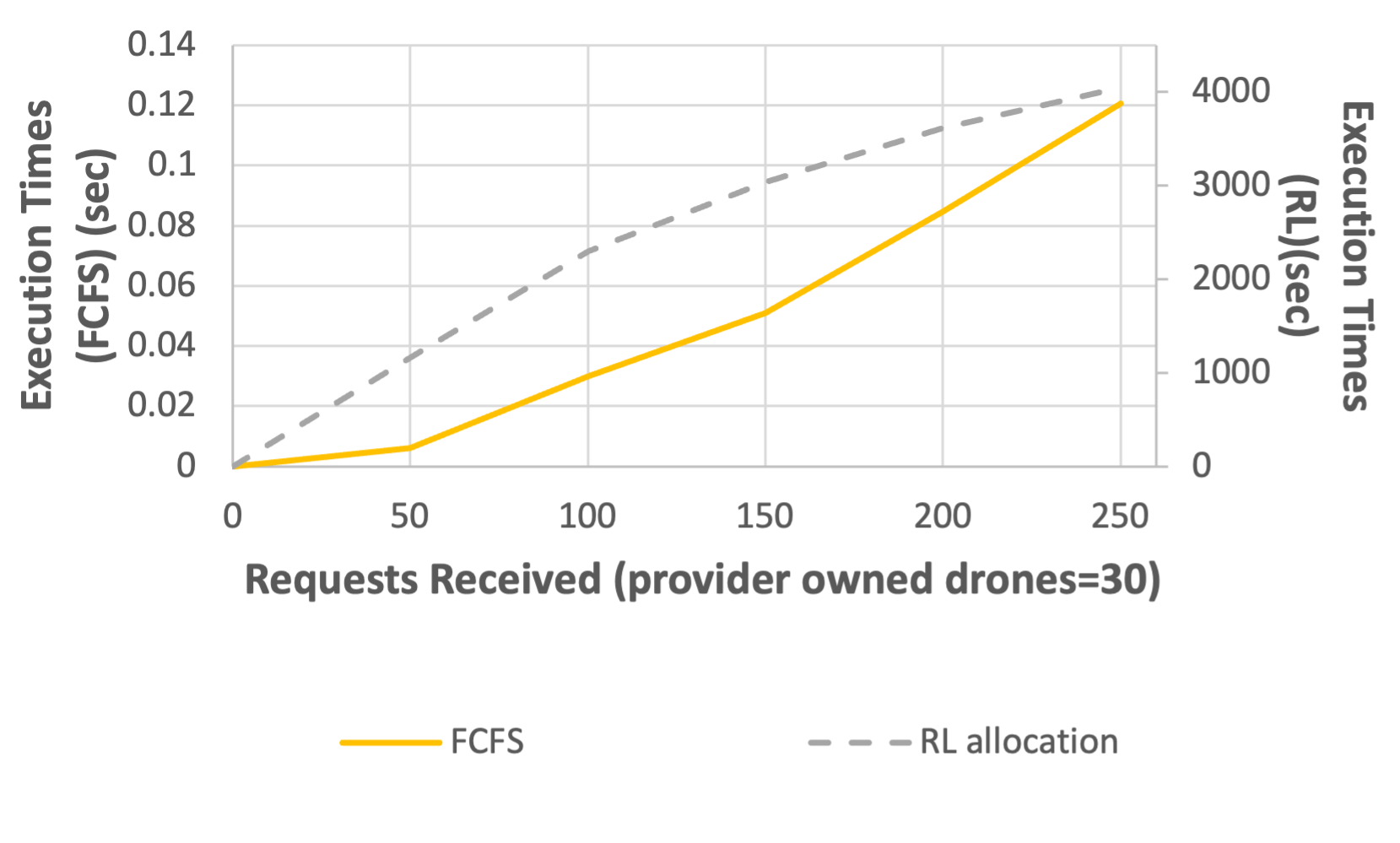}  
  \caption{Execution times}
  \label{fig:sub-seventh}
\end{subfigure}
\begin{subfigure}{.5\textwidth}
  \centering
  \includegraphics[width=.8\linewidth]{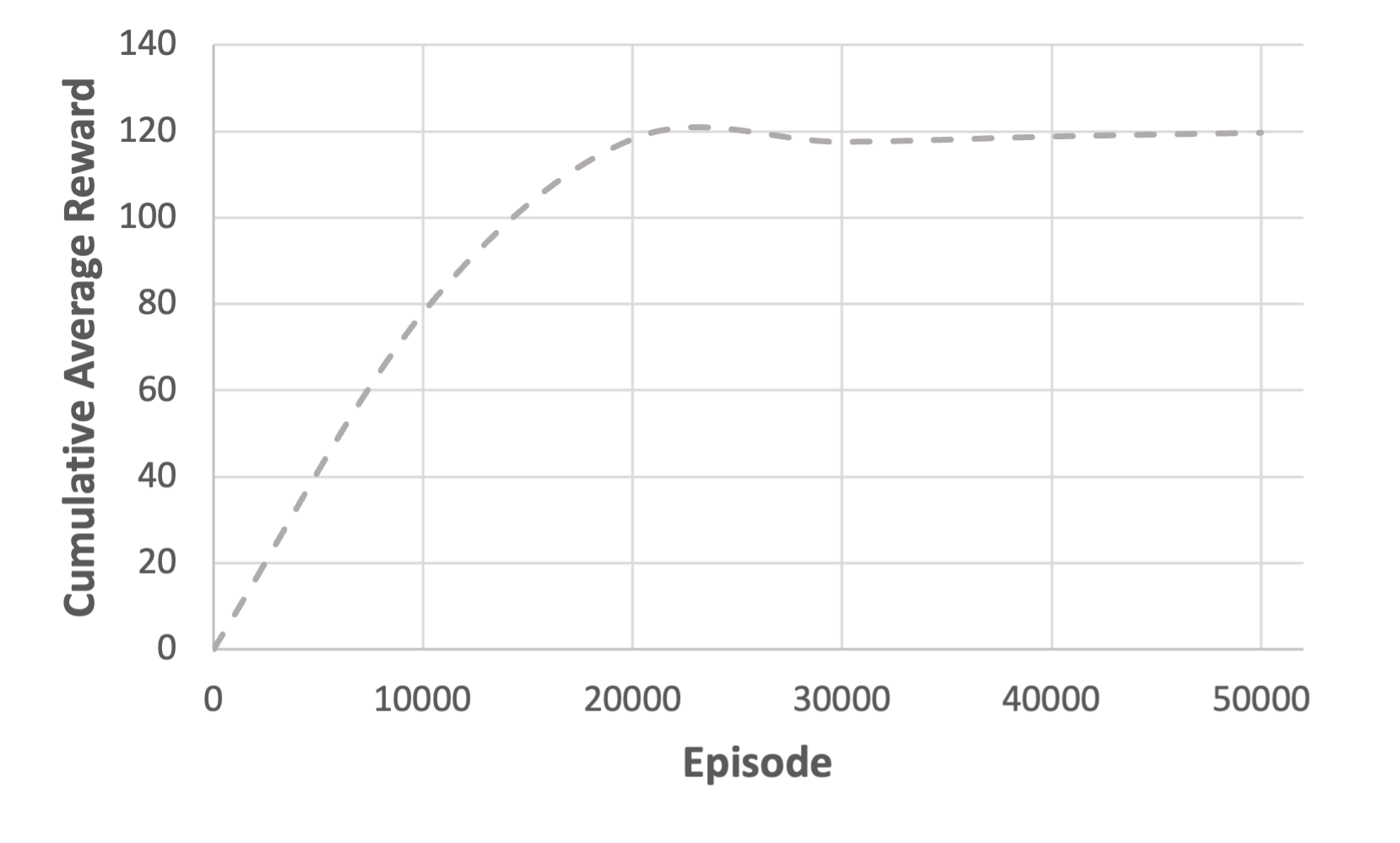}  
  \caption{RL rewards convergence}
  \label{fig:sub-eighth}
\end{subfigure}
\caption{Proposed Method Effectiveness}
\label{fig:fig}
\vspace{-0.0cm}
\end{figure}

In the first experiment, we study the effect of varying the number of received requests a day on the profit. We assume the provider owns a fixed set of 30 drones. As shown in Fig.\ref{fig:sub-first}, the RL allocation outperforms the FCFS. This is because of its ability to learn the optimal allocation and scheduling of the requests to maximize the profit. 
The FCFS is performing worse than RL because allocating services in an FCFS manner does not consider any order in terms of most profitable request and round trip times. Therefore, the non-optimal set of requests gets allocated at non-optimal time windows. 

The same behaviour is noted with varying the number of provider owned drones for a set of 50 requests as shown in Fig.\ref{fig:sub-second}. The RL allocation converges to the maximum possible profit earlier by serving all the 50 requests. This performance of the RL allocation method comes with the cost of execution. Fig.\ref{fig:sub-seventh} shows the execution times varying the number of requests received a day. The left  y-axis  represents  the  execution times of the FCFS. The right y-axis represents the execution time for the RL based algorithm. Since the number of state-action pairs in RL only increase in one dimension and converges at almost the 20000 episode (Fig.\ref{fig:sub-eighth}), the execution time does not increase significantly. We assume the requests are received in batch a day earlier, hence, the learning could occur overnight.

\section{Conclusion}
We proposed a provider-centric re-allocation of drone swarm services known as, Swarm-based Drone-as-a-Service (SDaaS). A congestion-aware SDaaS composition algorithm is proposed to compute the maximum delivery and round trip times a swarm may take to serve a request taking the constraints at intermediate nodes (limited recharging pads and congestion) in consideration. A reinforcement learning allocation  method was proposed with the goal of increasing the provider's profit. The efficiency of the proposed approach was evaluated in terms of profit maximization and execution time. Experimental results show the outperformance of the RL allocation approach to the baseline FCFS approach. In the future work, the problem could be expanded to cover multi-objectives, e.g. profit and time. In addition, we will consider heterogeneous swarms allocation to serve multiple requests and extend the work to deal with SDaaS failures.

\section*{Acknowledgment}
This research was partly made possible by DP160103595 and LE180100158 grants from the Australian Research Council. The statements made herein are solely the responsibility of the authors.

\bibliographystyle{splncs04}
\bibliography{samplepaper}
%




\end{document}